\documentclass[showpacs,showkeys, amsmath,amssymb, preprint, nofootinbib]{revtex4}
\usepackage{graphicx}
\usepackage{epsfig}
\usepackage{bm}
\usepackage{float}
\usepackage{fontenc}
\usepackage{graphicx}
\usepackage[dvips]{hyperref}
\usepackage{bm}
\usepackage{amsmath}
\def\beq{\begin{eqnarray}}

\def\eeq{\end{eqnarray}}

\begin{document}

\title[ISF]{Phantom solution in a non-linear Israel-Stewart theory}

\author{Miguel Cruz}
\email{miguelcruz02@uv.mx}
\address{Facultad de F\'\i sica, Universidad Veracruzana 91000, Xalapa, Veracruz, M\'exico}

\author{Norman Cruz}
\email{norman.cruz@usach.cl}
\address{Departamento de F\'\i sica, Universidad de Santiago de Chile, Casilla 307, Santiago, Chile}

\author{Samuel Lepe}
\email{samuel.lepe@pucv.cl}
\address{Instituto de F\'\i sica, Pontificia Universidad Cat\'olica de Valpara\'\i so, Casilla 4950, Valpara\'\i so, Chile}
\date{\today}

\begin{abstract}
In this paper we present a phantom solution with a big rip singularity in a non-linear regime of the Israel-Stewart formalism. In this framework it is possible to extend this causal formalism in order to describe accelerated expansion, where assumption of near equilibrium is no longer valid. We assume a flat universe filled with a single viscous fluid ruled by a barotropic EoS, $p=\omega \rho$, which can represent a late time accelerated phase of the cosmic evolution. The solution allows to cross the phantom divide without evoking an exotic matter fluid and the effective EoS parameter is always lesser than $-1$ and constant in time. 
\end{abstract}



\maketitle

\section{Introduction}

It is well known that along the evolution of our Universe are involved a great number of dissipative processes, however, there is not a definitive model to describe the overall dynamics of this dissipation since some explanations are based on speculative physics \cite{Maartens1}. In the search for a coherent model, was found that these dissipative processes can be well treated by employing a relativistic theory of bulk viscosity. Qualitatively, the bulk viscosity can be interpreted as a macroscopic consequence coming from the frictional effects in mixtures. The dark sector of the universe is the main material sector and is the one that presents more open questions about its real nature, in terms of what we know from our current theories. In this sense, one of the possible reasonable assumptions is to consider unified dark matter models, where the dark matter exhibits dissipative effects, which can lead the accelerated expansion that is associated to the dark energy component. For an inhomogeneous and isotropic universe only the bulk viscosity is present, and in general it is assumed that is ruled by a simple suitable expression in terms of the energy density of the dark fluid.  
A causal and stable theory of thermal phenomena in the presence of gravitational fields, was introduced by Israel and Stewart (IS) \cite{IS}, it provided a better description than Eckart and Landau and Lifshitz theories, but sharing with them the characteristic that only small deviations from equilibrium are assumed, then the transport equation is linear in the bulk viscous pressure; despite the improvements in the thermal description obtained with the IS theory, it was found in Ref. \cite{Maartens1} that for an inflating universe driven by viscosity it was necessary to consider a non-linear extension of the theory. This extension was done in \cite{Maartens2}, and the major aim of this work was to study inflationary solutions with no need of scalar fields, this new theory includes the IS theory in the linear regime ensuring the causality and stability, obeys the second law of thermodynamics, i.e., the positivity of the entropy production is ensured (see for instance the Ref. \cite{Cruz} with a discussion in this topic for the Horava-Lifshitz gravity in the holographic context) and naturally imposes an upper limit for the bulk viscous stress.\\ 
Additionally to the cosmological constant, the expansion of the Universe can be driven by scalar fields and the appropriate election of a potential function for the scalar field. Some of these scalar fields are candidates for Dark Energy (DE), such as quintessence and tachyonic fields, however, for these models the parameter $\omega$ of the equation state (EoS) $p = \omega \rho$ is restricted to $\omega > -1$. Nevertheless, more than a decade ago was pointed out in \cite{Obs1, Caldwell} that $\omega < -1$ is also compatible with most classical tests of cosmology, including the type 1a SNe data as well as the cosmic microwave background anisotropy and mass power spectrum. More recent observations do not ruled out this possibility \cite{Obs2, observations, observations1, observations2, Caldwell2}. In order to explore this new region it was neccesary the introduction of a new component given by a scalar field called {\it phantom} with ``wrong'' sign in its kinetic term, $\mathcal{L} = -\frac{1}{2}\partial_{\mu}\phi \partial^{\mu}\phi - V(\phi)$, such that $\rho + p = -\dot{\phi}^{2} < 0$. This scenario is not consistent with the dominant energy component (DEC), so an increasing energy density for the fluid filling the Universe occurs as the cosmic time evolves. These phantom fields have the drawback of experimenting vacuum instability \cite{instability}, so further investigations have been devoted to cross this phantom divide without instabilities. In the case of general k-essence models for a single scalar field which is described by an action with interaction with other energy components throughout kinetic couplings and higher derivatives, can not cross the phantom divide without gradient instabilities, singularities or ghosts. Another possibility, as suggested in \cite{Deffayet}, corresponds to the treatment of the scalar field as a velocity potential of an imperfect fluid, when the expansion around a perfect fluid is considered, the identification of some terms which correct the pressure in the manner of bulk viscosity can be performed.\\ 
In the case of imperfect fluids in the framework of General Relativity it is well known that the existence of dissipation in the cosmic components is a mechanism that allows violation of DEC condition \cite{Barrow}. Then, it is natural to explore the existence of phantom solutions in the causal thermodynamics approach of the IS theory. In this framework a phantom solution was recently found in Ref. \cite{CruzLepePena}. In this solution the Universe is filled with a single fluid with positive pressure and the viscosity drives the effective EoS to be of phantom type. So this solution can be interpreted as an indication of the possibility to cross the phantom divide via a dissipative mechanism. Nevertheless, a phantom solution do not satisfy the near equilibrium condition, given by
\begin{equation}
\left| \frac{\Pi }{\rho}\right| <<1,  
\label{eq:eq13}
\end{equation}
where $\Pi$ is the viscous pressure and $\rho$ the energy density of the fluid. This is because for an accelerated expansion the condition $\ddot{a}>0$ leads to
\begin{equation}
-\Pi > p +\frac{\rho}{3}. 
\label{Pieqa}
\end{equation}
Then, the inequality given by the Eq. (\ref{Pieqa}) implies that the viscous stress is greater than the equilibrium pressure $p$. The IS approach is valid in the near equilibrium regime, so, in order to include the phantom case it is necessary to consider a non-linear generalization of the causal linear thermodynamics of bulk viscosity where deviations from equilibrium are allowed. In this non-linear approach accelerated stable expansion has been reported for non-interacting two-fluid models where some of them presents viscosity \cite{Acquaviva1, Acquaviva2}. In this work we focus on a non-linear extension of the IS formalism to explore the possibility of a phantom solution, and how it can be characterized in terms of the parameters associated to the dissipative fluid involved.\\ 
The organisation of this article is as follows:  In Section II, by considering the non-linear extension of the IS theory made by Maartens and M\'endez \cite{Maartens2}, we explore the existence of a phantom solution for the dynamical equation of the non-linear bulk viscosity. This is done by using the following Ansatz
\begin{equation}
H\left( t\right) = A\left( t_{s}-t\right)^{-1}, \label{eq:Ansatz}
\end{equation}
which leads to a phantom solution for a late time FLRW flat universe filled with only one barotropic fluid with bulk viscosity, where $A$ is a constant and $ t_{s}$ a finite time in the future \cite{Cataldo2005}. 
In Section III we consider some thermodynamical implications of the phantom solution. In Section IV we write some general comments and the conclusions of the work.

\section{Phantom solution in the non-linear regime}
\label{sec:nlf}
For the obtention of a phantom solution in a non-linear extension of the causal IS formalism, we will follow the line of reasoning provided in Ref. \cite{Maartens2} where the universe has an unique component given by a fluid described by its pressure and density, denoted by $p$ and $\rho$ respectively. 
When scalar dissipative effects are considered, the large deviations from equilibrium arise from large bulk viscous stresses, this can be translated as $\left| \Pi \right| \geq p$, where $p$ is the local equilibrium pressure, and the non-equilibrium pressure is
\begin{equation}
p_{eff} = p + \Pi.
\label{eq:nep}
\end{equation}
For a covariant approach of causal thermodynamics of relativistic fluids, we make use of some hydrodynamic tensors together with the standard definition of some cosmological quantities. We define $n^{\alpha}$ as the particle number four-current and the entropy four-current $S^{\alpha}$. 
The non-negative entropy production rate or second law of thermodynamics $S^{\alpha}{}_{,\alpha} \geq 0$, must be guaranteed.
If we discard vector and tensor dissipation, we can express the entropy four-current as
\begin{equation}
S^{\alpha} = S_{eff}n^{\alpha},
\label{eq:s1}
\end{equation}
where $S_{eff}$ is the effective non-equilibrium specific entropy. In IS formalism, $S_{eff}$ is given by the expression
\begin{equation}
S_{eff} = S - \left(\frac{\tau}{2nT\zeta}\right)\Pi^{2},
\label{eq:seff}
\end{equation}
where $\zeta(\rho,n)$ is the bulk viscosity, $\tau(\rho,n)$ is the characteristic time for linear relaxational or transient effects and $T,n$ are the temperature and number density, respectively. $S$ and $T$ are local equilibrium variables which satisfy the Gibbs equation
\begin{equation}
TdS = (\rho + p)d\left(\frac{1}{n}\right)+\frac{1}{n}d\rho .
\label{eq:gibbs}
\end{equation}
In the IS framework, the viscous pressure obeys the following transport equation
\begin{equation}
\tau \dot{\Pi}+\left( 1+ \frac{1}{2}\tau \bigtriangleup \right)\Pi = -3\zeta H,
\end{equation}
where we have defined
\begin{equation}
\bigtriangleup := 3H+\frac{\dot{\tau}}{\tau}-\frac{\dot{\zeta}}{\zeta}-\frac{\dot{T}}{T}.
\end{equation}
On the other hand, the second law is satisfied by considering a linear relation between the thermodynamic ``force'' $\mathcal{X}$ and viscous pressure $\Pi$, i.e., $\Pi = -\zeta \mathcal{X}$ where
\begin{equation}
\mathcal{X} = 3H+\frac{\tau}{\zeta}\dot{\Pi}+\frac{\tau}{2\zeta}\Pi \bigtriangleup.
\label{eq:force}
\end{equation}
The non-linear effects can be introduced by considering the viscous pressure as \cite{Maartens2}
\begin{equation}
\Pi = -\frac{\zeta \mathcal{X}}{1+\tau_{*}\mathcal{X}},
\label{eq:nlp}
\end{equation}
where $\tau_{*} \geq 0$ is the characteristic time for non-linear effects and is defined as 
\begin{equation}
\tau_{*}=k^{2}\tau,
\label{eq:new2}
\end{equation}
being $k$ a constant. In the case $k=0$ we obtain the linear IS theory. The linear relaxation time and the bulk viscosity are related as
\begin{equation}
\tau = \frac{\zeta}{v^{2}(\rho + p)},
\label{eq:time}
\end{equation}
being $v^{2}$ the dissipative contribution to the speed of sound, $V$. The causality condition demands $V^{2} =v^{2}+c^{2}_{s} \leq 1$, where $c^{2}_{s} = (\partial p/\partial \rho)_{S}$ is the adiabatic contribution. In the barotropic case, $p = \omega \rho$, we obtain $c^{2}_{s} = \omega$ then $v^{2} \leq 1-\omega$. 
In the non-linear case the second law of thermodynamics holds by establishing an upper bound on the bulk stress, using the Eqs. (\ref{eq:force}) and (\ref{eq:nlp}) we can express
\begin{equation}
S^{\alpha}{}_{,\alpha}=n\dot{S}_{eff}=\frac{\Pi^{2}}{T\zeta}\left(1+\frac{\tau_{*}}{\zeta}\Pi \right)^{-1},\ \ \ \mbox{then} \ \ \ S^{\alpha}{}_{,\alpha} > 0 \ \ \mbox{when}\ \ -\Pi < \frac{\zeta}{\tau_{*}},
\label{eq:2nd}
\end{equation}
as $-\Pi$ approaches to $\zeta/\tau_{*}$ the entropy exhibits a singular behavior, then, the introduction of non-linear effects imposes a restriction on the viscous pressure $\Pi$.  
From now on, we consider a flat FLRW geometry, so in this case the equations of motion are the following ($8\pi G = c = 1$ units will be used in this work) 
\begin{eqnarray}
H^{2} &=& \frac{1}{3}\rho,\label{eq:eom1}\\
\dot{H}+H^{2} &=& -\frac{1}{6}\left[\rho +  3\left(p + \Pi \right) \right].\label{eq:eom2}
\end{eqnarray}
For the barotropic EoS, $p = \omega \rho$, where $0\leq \omega < 1$ in order to satisfy the causality condition mentioned before, for this case the continuity equation is given by 
\begin{equation}
\dot{\rho}+3H\left[\left(1+\omega \right)\rho + \Pi \right] = 0.
\label{eq:cont}
\end{equation} 
By means of the integrability Gibbs condition we can find for the temperature
\begin{equation}
T \propto \rho^{\omega/(\omega + 1)}.
\label{eq:temp}
\end{equation}
Generally, the bulk viscosity for a FLRW spacetime can be taken in the following simple form $\zeta =\xi_{0}\rho^{s}$, where $\xi_{0}$ is a constant parameter obeying the condition $\xi_{0}\geq 0$, which is necessary for an expanding universe. In the framework of the Eckart formalism $\xi_{0}$ has been constrained using observational data such as SNe Ia, CMB, BAO and Gamma Ray Bursts (GRBs) calibration \cite{Yoe, Nucamendi1, Nucamendi2, Nucamendi3, Montiel}. Using the equations of motion (\ref{eq:eom1})-(\ref{eq:eom2}), we have
\begin{equation}
\zeta = 3^{s}\xi_{0}H^{2s}.
\label{eq:new3}
\end{equation} 
For $s=1/2$ one gets $\zeta = \sqrt{3}\xi_{0}H$. On the other hand, using the barotropic EoS $p=\omega \rho$, and the expressions (\ref{eq:eom1})-(\ref{eq:eom2}) we also can write an explicit expression for the dissipation term
\begin{equation}
\Pi = -2\dot{H}-3(\omega + 1) H^{2}.
\label{eq:Pi}
\end{equation}
In this case the fundamental dynamical equation for non-linear bulk viscosity in a flat universe can be written as \cite{Maartens2}, 
\begin{align}
& \left[1-\frac{k^{2}}{v^{2}}-\left(\frac{2k^{2}}{3(\omega + 1) v^{2}} \right)\frac{\dot{H}}{H^{2}} \right]\left\lbrace \ddot{H}+3H\dot{H}+\left(\frac{1-2(\omega + 1)}{(\omega + 1)} \right)\frac{\dot{H}^{2}}{H}+\frac{9}{4}(\omega + 1) H^{3} \right\rbrace \nonumber \\
& + \frac{3(\omega + 1) v^{2}}{2\sqrt{3}\xi_{0}}\left[1+\left(\frac{\sqrt{3}\xi_{0} k^{2}}{(\omega + 1) v^{2}}\right)H^{2s-1} \right]H^{2(1-s)}(2\dot{H}+3(\omega + 1) H ^{2})-\frac{9}{2}(\omega + 1) v^{2}H^{3} = 0.
\label{eq:brigida}
\end{align}
We will inspect for a solution of Eq. (\ref{eq:brigida}) of the form given in Eq. (\ref{eq:Ansatz}). This solution represents a big rip singularity \cite{Sergei} and has already found in causal viscous schemes \cite{CruzLepePena, Norman}.
 
\subsection{Phantom solution}
\label{sec:PS}
\begin{figure}
\centering
\includegraphics[scale=0.7]{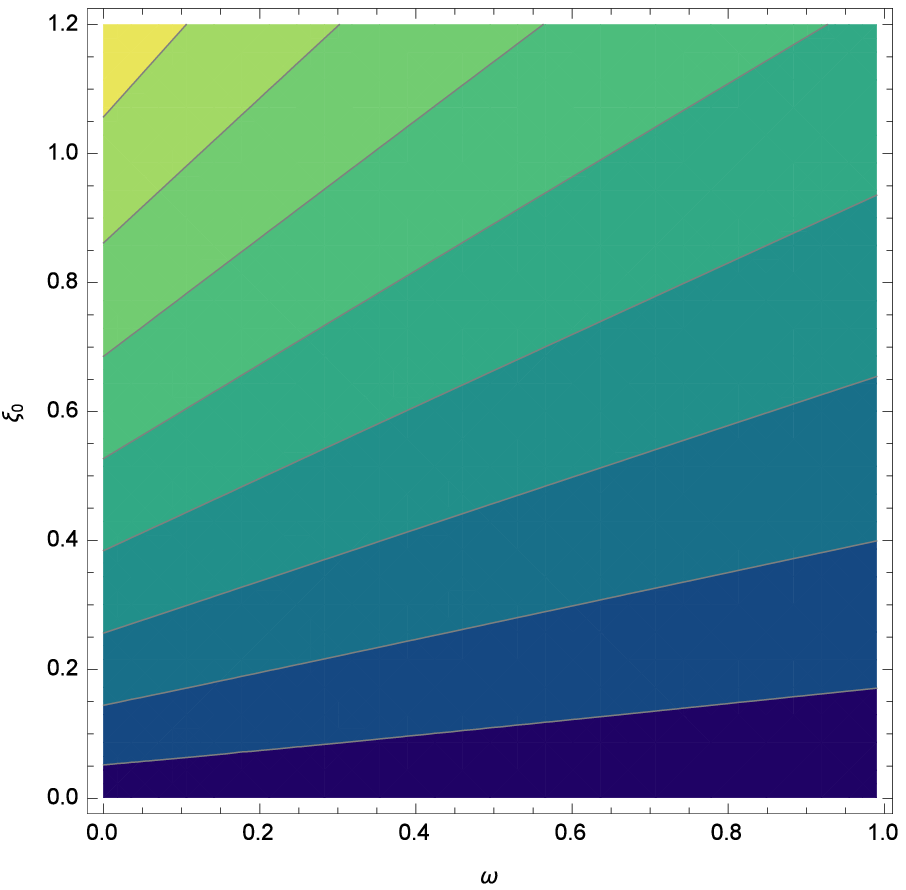}
\includegraphics[scale=1]{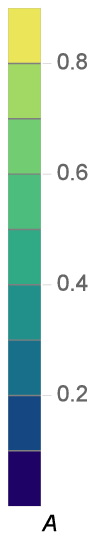}
\includegraphics[scale=0.7]{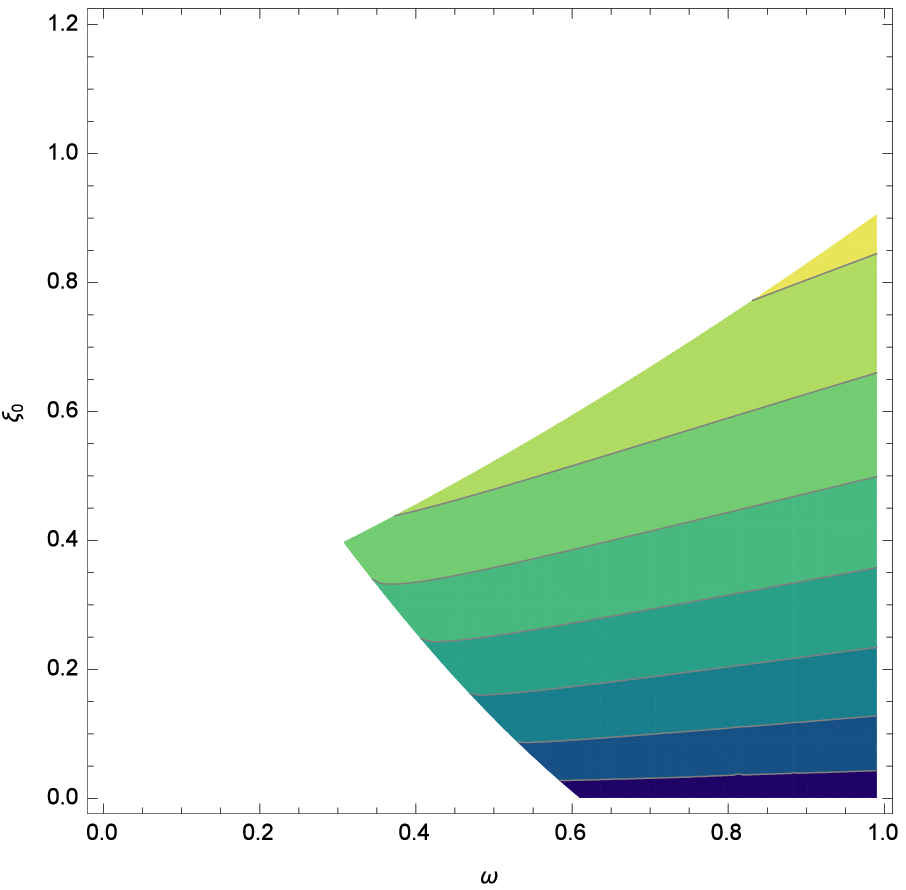}
\includegraphics[scale=1]{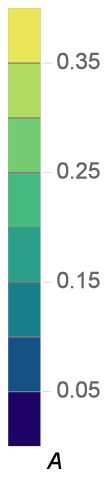}
\caption{Positive solution for the variable $A$ in the space of parameters $(\omega, \xi_{0})$. In both panels the value of $v$ is the same, however, for the right panel the value of $k$ is close to the value of $v$.} 
\label{fig:possol}
\end{figure}

In this section we focus on the possibility of finding a phantom solution for the dynamical equation of the non-linear bulk viscosity (\ref{eq:brigida}). It is straightforward to see that this equation becomes easier to manage in the special case where the bulk viscosity coefficient takes the form $\zeta (\rho) \sim \rho^{1/2}$. Previous works have found solutions which presents a big rip singularity for the above election in the IS causal linear thermodynamics formalism. We will apply the Ansatz given by Eq. (\ref{eq:Ansatz}), after substitution in (\ref{eq:brigida}) and direct computation we are left with the following fourth order algebraic equation for $A$
\begin{equation}
C_{4}A^{4}+C_{3}A^{3}+C_{2}A^{2}+C_{1}A-C_{0} = 0,
\label{eq:fourth}
\end{equation}
which must have at least one real and positive solution. The coefficients involved in the last equation are defined as follows
\begin{eqnarray}
C_{4} &=& 3a_{4}(1+\omega)(1+a_{5}),\label{eq:coeff}\\
C_{3} &=& \frac{9}{2}(1+\omega)\left(\frac{a_{1}}{2}-v^{2}\right),\\
C_{2} &=& 2a_{4}(1+a_{5})+3a_{1}-\frac{9}{4}a_{2}(1+\omega),\\
C_{1} &=& a_{1}(2+a_{3})-3a_{2},\\
C_{0} &=& a_{2}(2+a_{3}),
\end{eqnarray}
and
\begin{eqnarray}
a_{1} &=& 1-\frac{k^{2}}{v^{2}},\\
a_{2} &=& \frac{2k^{2}}{3v^{2}(1+\omega)},\\
a_{3} &=& -\left(\frac{1+2\omega}{1+\omega} \right),\\
a_{4} &=& \frac{\sqrt{3}(1+\omega)v^{2}}{2\xi_{0}},\\
a_{5} &=& \frac{k^{2}}{v^{2}}\left(\frac{\sqrt{3}\xi_{0}}{1+\omega} \right)\label{eq:par}.
\end{eqnarray}
For the parameters involved on the second set of equations we will restrict ourselves to the following constraints: $k^{2} \leq v^{2}$, $0 < \xi_{0} < 1$ and $0 \leq \omega < 1$. Using these conditions together with Eqs. (\ref{eq:coeff})-(\ref{eq:par}) and following the procedure exposed in Ref. \cite{Abramowitz}, we solve the Eq. (\ref{eq:fourth}) for the variable $A$ (more details can be seen in Appendix (\ref{sec:appA})). If the constraints mentioned above are satisfied, we can find that Eq. (\ref{eq:fourth}) possess an unique positive solution. The behavior of the positive solution is depicted in Fig. (\ref{fig:possol}) in the space of parameters $(\omega, \xi_{0})$.\\

\begin{figure}
\centering
\includegraphics[scale=0.7]{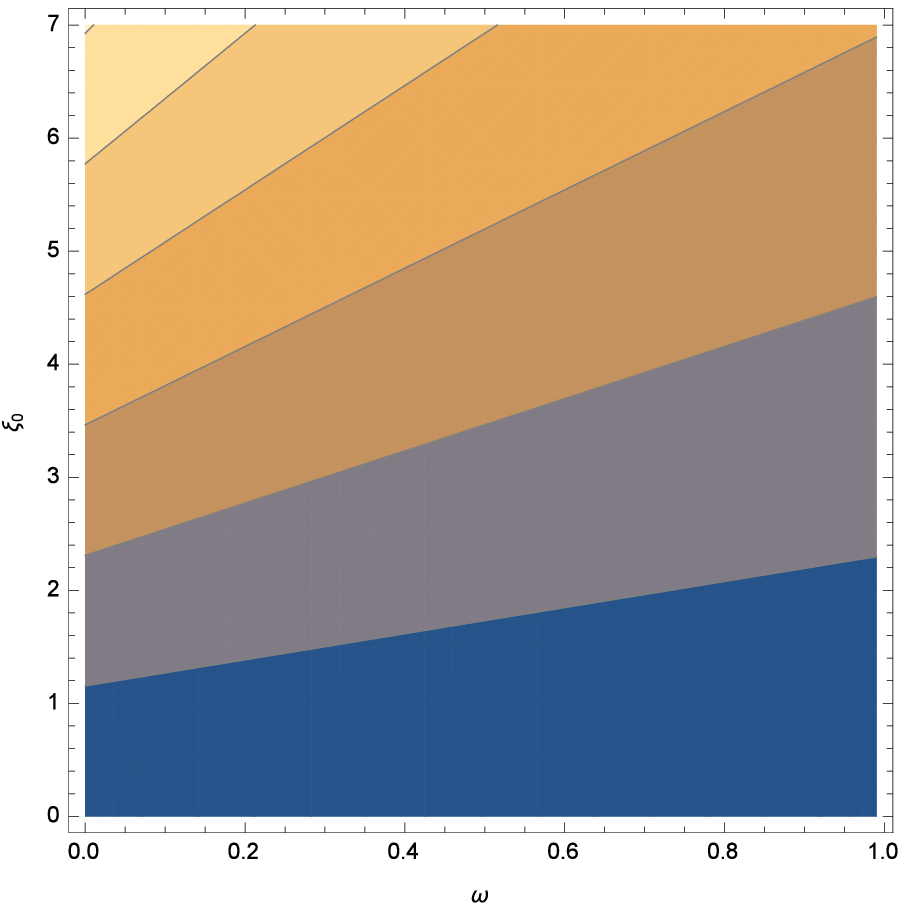}
\includegraphics[scale=1]{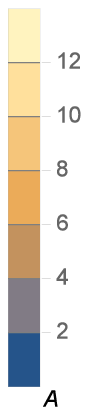}
\caption{Positive solution for the variable $A$ considering the third case of the inequality $k^{2} \leq v^{2}$, i.e., the condition $v \gg k$ for a fixed value of the parameter $k$.}
\label{fig:possol1}
\end{figure}

Some remarks are in order: we discuss the threefold situation for the inequality $k^{2} \leq v^{2}$. (1) We consider a fixed value for the parameter $v$, as the parameter $k$ becomes smaller, the positive solution $A$ can be extended to the region where $\xi_{0} > 1$, this can be visualized in the left panel of Fig. (\ref{fig:possol}), as $k$ approaches to $v$, the region in the space of parameters $(\omega, \xi_{0})$ where the positive solution exists, becomes smaller. (2) For the case $k^{2} = v^{2}$ there is not positive solution for the variable $A$.
The IS theory is recovered in the limiting case $k=0$ and we also have a positive solution for $A$. (3) We fix the value of the parameter $k$, when $v \gg k$ we have an unique positive solution for $A$, the behavior of this solution is plotted in Fig. (\ref{fig:possol1}).

\section{Thermodynamical Properties of Phantom Solution}
\begin{figure}
\centering
\includegraphics[scale=0.9]{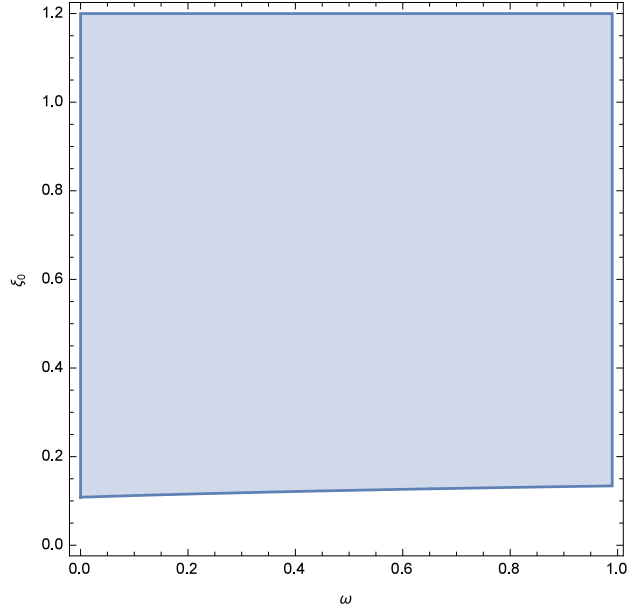}
\caption{In the space of parameters $(\xi_{0},\omega)$, the condition given by the expression (\ref{eq:ineq}) is satisfied by the solution shown on the left panel of Fig. (\ref{fig:possol}) only within the blue region. In this region a positive entropy production is guaranteed.}
\label{fig:ineq1}
\end{figure}

\begin{figure}
\centering
\includegraphics[scale=0.9]{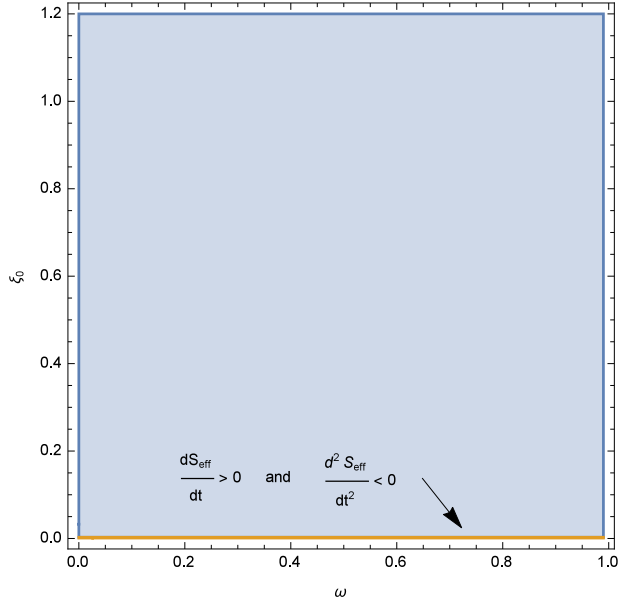}
\caption{Positive entropy production and convexity conditions for the solution $A$ obtained with the condition $k \ll v$.}
\label{fig:overlap1}
\end{figure}

\begin{figure}
\centering
\includegraphics[scale=0.9]{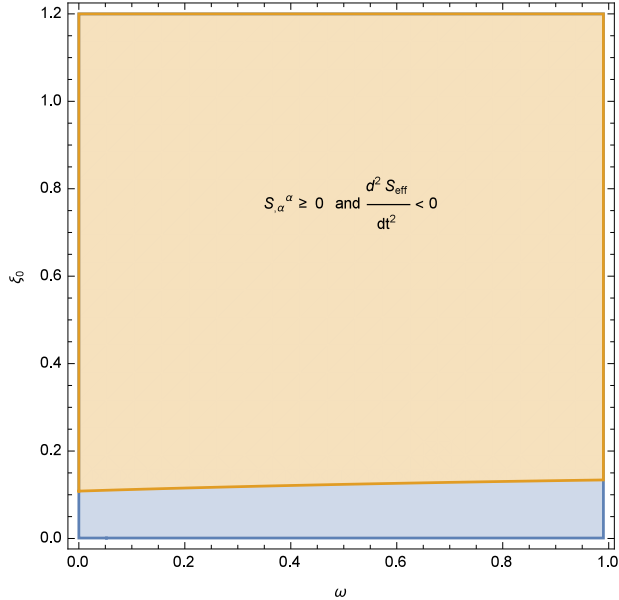}
\caption{Positive entropy production and convexity conditions.}
\label{fig:overlap}
\end{figure}

\begin{figure}
\centering
\includegraphics[scale=0.7]{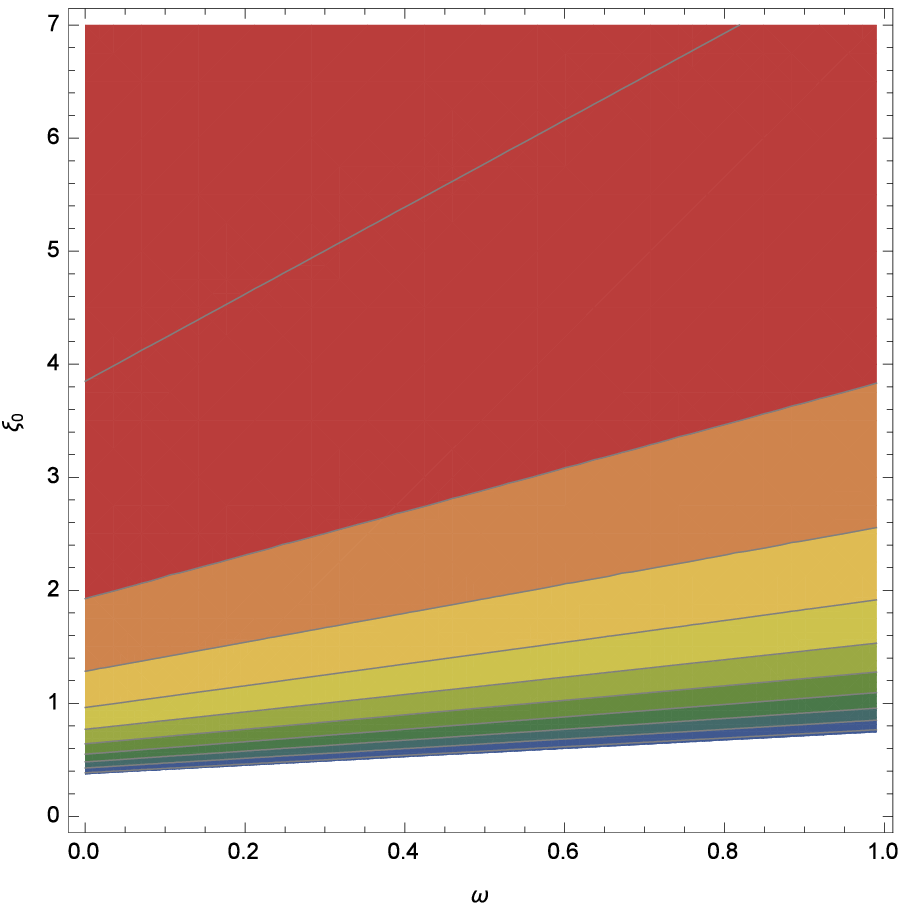}
\includegraphics[scale=1]{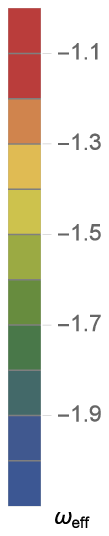}
\caption{Effective $\omega_{eff}$ in the parameter space $(\omega, \xi_{0})$.}
\label{fig:omega}
\end{figure}

If we consider the de Sitter case, i.e., $H=H_{0} = \mbox{constant}$ in Eq. (\ref{eq:brigida}) it is straightforward to obtain an expression for $H_{0}$ which is the same obtained in Ref. \cite{Maartens2}\footnote{From Eq. (\ref{eq:brigida}) we can obtain
\begin{equation*}
H_{0}^{1-2s} = \frac{\sqrt{3}\xi_{0}}{2(\omega + 1)v^{4}}\left(2v^{2}-1 \right)\left(v^{2}-k^{2} \right),
\end{equation*}
where $H_{0}\geq 0$ in the limit $k \leq v$ and $v^{2} > 1/2$.}. This solution can describe an inflationary process, with an entropy and effective specific entropy positive and both exhibiting an exponential growth. By means of Ansatz (\ref{eq:Ansatz}) used to find the phantom solution in the non-linear framework, we can evaluate explicitly the number density of particles, the scale factor and the viscous pressure. The number density of particles can be found from the conservation law in a FLRW geometry
\begin{equation}
\dot{n} + 3Hn = 0,
\label{eq:continuity}
\end{equation}
then, 
\begin{equation}
n(t) = n_{0}\left(\frac{t_{s}-t}{t_{s}-t_{0}}\right)^{3A},
\end{equation}
and for the scale factor one gets
\begin{equation}
a(t) = a_{0}\left(\frac{t_{s}-t}{t_{s}-t_{0}} \right)^{-A}.
\end{equation}
When $t_{s} = t$, the size of the universe becomes infinite and the density of particles goes to zero. The viscous pressure is evaluated from Eq. (\ref{eq:Pi}), yielding
\begin{equation}
\Pi(t) = -A(t_{s}-t)^{-2}\left[3A(\omega + 1)+2\right].
\label{eq:new4}
\end{equation}
The condition written in Eq. (\ref{eq:2nd}) implies, using the Eqs. (\ref{eq:time}), (\ref{eq:new2}), (\ref{eq:new3}) and (\ref{eq:new4}), a limit on the parameter $k$
\begin{equation}
k^{2} \leq \frac{v^{2}}{1+\frac{2}{3A(\omega +1)}},
\label{eq:ineq} 
\end{equation}
where $A$ is the solution found previously. Then, this last condition must be guaranteed for all the values given to both parameters $k$ and $v$ in the solution $A$. Despite we have a positive solution for $A$ in the three cases considered before, not all values of $k$ and $v$ satisfy the condition (\ref{eq:ineq}). The solution shown in the right panel of Fig. (\ref{fig:possol}) is discarded by the limit on $k$. The solution $A$ shown in Fig. (\ref{fig:possol1}), satisfies the inequality (\ref{eq:ineq}). Finally, for the solution shown in the left panel of Fig. (\ref{fig:possol}) we must exclude a region of the space of parameters $(\xi_{0},\omega)$ in order to satisfy the condition (\ref{eq:ineq}), this can be seen in Fig. (\ref{fig:ineq1}). 
The entropy production expression
\begin{equation}
\frac{dS}{dt} = -\frac{3H\Pi}{nT},
\label{eq:s}
\end{equation}
can be calculated from the Gibbs equation (\ref{eq:gibbs}) and the Eqs. (\ref{eq:cont}), (\ref{eq:continuity}). Using this last result together with Eqs. (\ref{eq:seff}) and (\ref{eq:time}), we can compute the first derivative of the effective specific entropy\footnote{In the non-linear regime we can consider the effective specific entropy which is more suitable far from equilibrium \cite{Maartens2}.}, obtaining
\begin{eqnarray}
\frac{dS_{eff}}{dt} &=& \frac{(3A^{2})^{1/(\omega +1)}(t_{s}-t_{0})^{3A}\left[2+3A(\omega + 1) \right]}{2v^{2}n_{0}T_{0}[3A(\omega+1)]^{2}}\left[3A(\omega +1)\left\lbrace 3A(\omega + 1)(2v^{2}-1)-4 \right\rbrace - 4\right]\times \nonumber \\
& \times & (t_{s}-t)^{-\frac{(3A+1)(\omega + 1)+2}{(\omega +1)}},
\label{eq:first}
\end{eqnarray}
and the second derivative can be written as follows
\begin{eqnarray}
\frac{d^{2}S_{eff}}{dt^{2}} &=& \frac{(3A^{2})^{1/(\omega +1)}(t_{s}-t_{0})^{3A}\left[2+3A(\omega + 1) \right]\left[(\omega + 3)+3A(\omega +1) \right]}{2v^{2}n_{0}T_{0}\left(\omega +1\right)[3A(\omega+1)]^{2}}\times \nonumber \\
& \times & \left[3A(\omega +1)\left\lbrace 3A(\omega + 1)(2v^{2}-1)-4 \right\rbrace - 4\right](t_{s}-t)^{-\frac{(3A+1)(\omega + 1)+(\omega +3)}{(\omega +1)}}.
\label{eq:second}
\end{eqnarray}
For thermodynamical consistency two conditions must be satisfied by the entropy function: $dS_{eff}/dt > 0 \longleftrightarrow v^{2} > \Theta/2$ and $d^{2}S_{eff}/dt^{2} < 0 \longleftrightarrow v^{2} < \Theta/2$, where, according to Eqs. (\ref{eq:first}) or (\ref{eq:second}),
\begin{equation}
\Theta = 1+\frac{4}{3A(\omega +1)}\left[1+\frac{1}{3A(\omega +1)} \right] > 1,
\end{equation}
and, obviously, only one condition for $v^{2}$ can be satisfied in accordance to specific values for the parameters involved in the model. However, by taking into account the condition $k \ll v$, which is one of the cases discussed before for the solutions $A$, the conditions  $dS_{eff}/dt > 0$ and $d^{2}S_{eff}/dt^{2} < 0$ can be satisfied simultaneously, but only around small values of $\xi_{0}$ as can be seen in Fig. (\ref{fig:overlap1}). This is not longer valid for the other cases of the solution $A$. 
On the other hand, as a second example, since we are considering non-linear effects and a phantom solution, the second law of thermodynamics (\ref{eq:2nd}) holds if the condition (\ref{eq:ineq}) is satisfied, if we consider one of the solutions discussed previously\footnote{The solution for $A$ shown on the left panel of Fig. (\ref{fig:possol}).}, as can be seen in Fig. (\ref{fig:overlap}) within the space of parameters $(\omega, \xi_{0})$ exists a larger overlapped region where the positive entropy production given by the condition (\ref{eq:ineq}) and the entropy convexity, $d^{2}S_{eff}/dt^{2} < 0$, are guaranteed. As we showed, when the phantom solution is considered in a non-linear IS theory, we have that the thermodynamical consistency demanded for the evolution of natural process only apply for one of the cases where the phantom solution is admitted. Nevertheless, we have pointed out that as we approach to the singularity  the deviations from equilibrium goes to infinity, which can be seen from the divergences of the effective specific entropy, its derivatives and the viscous pressure. This is a strong indication that even the non-linear generalization of the causal thermodynamics can not necessarily be valid in this regime.
For definiteness, in Eq. (\ref{eq:temp}) the proportionality constant will be established as $T_{0}$, then
\begin{equation}
T(t) = T_{0}\left[3A^{2}\left(t_{s}-t\right)^{-2} \right]^{\omega/(\omega+1)},
\end{equation}
using these expressions and after integration of the Eq. (\ref{eq:s}) we obtain
\begin{equation}
S = S_{0}+\frac{(3A^{2})^{1/(\omega+1)}(t_{s}-t_{0})^{3A}(\omega + 1)}{n_{0}T_{0}}(t_{s}-t)^{-\frac{3A(\omega+1)+2}{\omega + 1}},
\end{equation}
where $S_{0}$ is an integration constant, for the effective specific entropy we have
\begin{eqnarray}
S_{eff} &=& S_{0}+\frac{(3A^{2})^{-\omega/(\omega+1)}(t_{s}-t_{0})^{3A}}{6n_{0}T_{0}v^{2}(\omega+1)}\left[9A^{2}(\omega + 1)^{2}\left(2v^{2}-1 \right)-12A(\omega +1)-4 \right]\times \nonumber \\
& \times & (t_{s}-t)^{-\frac{3A(\omega+1)+2}{\omega + 1}}.
\end{eqnarray}
Some remarks are in order. For consistency at thermodynamical level the positivity of the effective specific entropy must be guaranteed in the form $v^{2}> \Theta/2$. On the other hand, if $t_{s} -t \gg 1$, the entropy tends to a constant. In the case $t_{s} = t$ we have a singular behavior for the entropy. Using the expressions (\ref{eq:Ansatz}), (\ref{eq:nep}) and (\ref{eq:Pi}) we can find the effective EoS of the phantom solution, characterized by a parameter $\omega_{eff}$, which take the form 
\begin{equation}
\omega_{eff} = \omega + \frac{\Pi(t)}{\rho(t)} = -1-\frac{2}{3A},
\label{eq:eff}
\end{equation}
where $A$ can take the values for the case $v \gg k$, which is showed in Fig. (\ref{fig:possol1}). In Fig. (\ref{fig:omega}) we can observe the behavior of the function expressed in Eq. (\ref{eq:eff}) in the space of parameters $(\omega, \xi_{0})$. In the case of measurements in the framework of $\Lambda CDM$ model, the value $\omega_{eff} \sim -1.1$ has been reported in \cite{observations}, for the EoS associated to the dark energy component. In our model we do not have a dark energy component and then $\Omega_{m} =1$, in the case of a flat universe. This means that in the model studied the simplification of neglecting other matter components with negative pressure, as in the $\Lambda CDM$, is the price to pay to obtain a phantom solution in the non-linear thermodynamics approach of the IS theory. In our model values around $\omega_{eff} \sim -1.1$ can be obtained for large values of $\xi_{0}$ for $0 \leq \omega < 1$.   

\section{Final remarks}
This article was devoted to study a phantom solution of the type $H(t) = A(t_{s}-t)^{-1}$ within the framework of the causal non-linear extension of the IS formalism. This was done for a flat-FLRW universe filled with one single fluid characterized by a barotropic EoS and the bulk viscosity ruled by the simple form $\zeta =\xi_{0}\rho^{s}$. We have found that this phantom solution under the condition $v \gg k$ is allowed and for a fixed value for the parameter $v$, a positive solution for $A$ can be extended to the region where $\xi_{0} > 1$.\\
 
At thermodynamical level we observed that in the non-linear extension of the IS formalism we must always make sure that condition (\ref{eq:ineq}) is satisfied, in order to have positive entropy production. The constraint $v \gg k$ allows to fulfill both conditions: $dS_{eff}/dt > 0$ and $d^{2}S_{eff}/dt^{2} < 0$. Nevertheless, since we are far from equilibrium as we approach to the singularity, the condition $d^{2}S_{eff}/dt^{2} < 0$ can not be demanded as a test for a well thermodynamic behavior. The divergences of the thermodynamical parameters, like effective specific entropy, its derivatives and the viscous pressure at the singularity, leads to huge deviations from equilibrium. In consequence, even in the framework of the non-linear generalization of the causal thermodynamics, the phantom solution obtained deserves further investigation to elucidate fully its thermodynamical behavior.\\
 
Finally, we have a phantom solution with an effective time independent EoS, which is always lesser that $-1$. This phantom behavior have been obtained for a flat universe with only one matter component, which has positive pressure and bulk viscosity. Then in the framework of the non-linear thermodynamics approach of the IS theory it is possible to obtain phantom behavior without invoking phantom fields. If we assume that our model could be a suitable first approximation to obtain phantom behavior with dissipative normal matter, then to obtain values around $-1.1$ for the parameter $\omega_{eff}$, as reported using observational data, are possible only for large values of $\xi_{0}$, in the range $0 \leq \omega < 1$. 

\section*{Acknowledgments}
N. C. acknowledges the hospitality of the Facultad de F\'\i sica of
Universidad Veracruzana, M\'exico, where part of this work was done. This
work has been supported by CONACyT-M\'exico through the grant
Repatriaciones, 2015, cuarta fase and SNI (M. C.) and N. C. and S. L.
acknowledge the support of following Chilean Institutions: CONICYT, through
the Fondecyt grant No. 1140238 (N. C.) and Pontificia Universidad Cat\'olica de 
Valpara\'\i so, Chile, through the VRIEA-DI-PUCV grant No. 039.351/2016 (S. L.).

\appendix
\section{Solution for the fourth order algebraic equation}
\label{sec:appA}
We write the Eq. (\ref{eq:fourth}) as follows
\begin{equation}
A^{4}+\mathbf{C}_{3}A^{3}+\mathbf{C}_{2}A^{2}+\mathbf{C}_{1}A-\mathbf{C}_{0} = 0,
\end{equation}
where we have defined $\mathbf{C}_{i} = C_{i}/C_{4}$ being $i = 0, 1, 2, 3$. With the coefficients of the previous equation, we construct the cubic equation
\begin{equation}
u^{3}-\mathbf{C}_{2}u^{2}+(\mathbf{C}_{1}\mathbf{C}_{3}+4\mathbf{C}_{0})u-(\mathbf{C}^{2}_{1}-\mathbf{C}_{0}\mathbf{C}^{2}_{3}+4\mathbf{C}_{0}\mathbf{C}_{2}) = 0,
\end{equation}
which must have at least one real root, under the standard Liouville change of variable $u = q + \mathbf{C}_{2}/3$, this equation can be written in the form of an incomplete cubic equation\footnote{In general the incomplete cubic equation has the form $q^{3}+mq+n = 0$.}, i.e.,
\begin{equation}
q^{3}-\frac{1}{3}\left(\mathbf{C}^{2}_{2}-3\mathbf{C}_{1}\mathbf{C}_{3}-12\mathbf{C}_{0} \right)q +\frac{1}{3}\left(\mathbf{C}_{1}\mathbf{C}_{2}\mathbf{C}_{3}-\frac{2}{9}\mathbf{C}^{3}_{2}-8\mathbf{C}_{0}\mathbf{C}_{2}-\mathbf{C}^{2}_{1}+\mathbf{C}_{0}\mathbf{C}^{2}_{3} \right) = 0.
\end{equation}
This equation can be solved directly by using the Cardano formulae 
\begin{eqnarray}
q_{1} &=& \mathcal{A}+\mathcal{B},\\
q_{2,3} &=& -\frac{1}{2}\left(\mathcal{A}+\mathcal{B} \right) \pm i\frac{\sqrt{3}}{2}\left(\mathcal{A}-\mathcal{B} \right),
\end{eqnarray}
where $\mathcal{A} = \left(-n/2+\sqrt{D}\right)^{1/3}$ and $\mathcal{B}= \left(-n/2-\sqrt{D}\right)^{1/3}$, being $D$ the discriminant, $D = (m/3)^{3}+(n/2)^{2}$. In dependence on the value of the discriminant we can have the following situations: (i) $D > 0$ one real root, two complex (ii) $D<0$ three real roots (iii) $D=0$ three real roots, two are degenerate. Using the constraints defined on the parameters, we are left with the (i) case, $D > 0$. Once we solve for $q$ by means of the Liouville change of variable we can recover $u$. Then, with this real root we construct the following quadratic equations
\begin{equation}
v^{2}+\left[\frac{\mathbf{C}_{3}}{2}\mp \left(\frac{\mathbf{C}^{2}_{3}}{4}+u-\mathbf{C}_{2}\right)^{1/2}\right]v+\frac{u}{2}\mp \left[\left(\frac{u}{2}\right)^{2}+\mathbf{C}_{0}\right]^{1/2} = 0,
\end{equation}
where the positive solution can be obtained within the space of parameters $(\xi_{0},\omega)$ as shown before.

\end{document}